\documentstyle[prb,aps]{revtex}
\input epsf

\begin{document}
\draft
\wideabs{
\title{Reflection and transmission of waves
in surface-disordered waveguides}
\author{J. A. S{\'a}nchez-Gil}
\address{Instituto de Estructura de la Materia, 
Consejo Superior de Investigaciones Cient{\'\i}ficas,
Serrano 121, 28006 Madrid, Spain}
\author{V. Freilikher}
\address{The Jack and Pearl Resnick Institute of Advanced Technology,
Department of Physics \\ Bar-Ilan University, Ramat-Gan 52900, Israel}
\author{A. A. Maradudin}
\address{Department of Physics and Astronomy and Institute for Surface
and Interface Science \\ University of California, Irvine, CA 92697}
\author{I. V. Yurkevich}
\address{School of Physics and Space Research, University of Birmingham,
Edgbaston, Birmingham B15 2TT, UK}

\date{\today}
\maketitle

\begin{abstract}
The reflection and transmission amplitudes of waves in disordered
multimode waveguides are studied by means of numerical simulations
based on the invariant embedding equations. 
In particular, we analyze the influence of  surface-type disorder on
the behavior of the ensemble average and fluctuations of the reflection
and transmission coefficients, reflectance, transmittance, and
conductance. Our results show anomalous effects stemming from the
combination of mode dispersion and rough surface scattering: For a given
waveguide length, the larger the mode transverse momentum is, the 
more strongly is the mode  scattered. These effects manifest themselves in the
mode selectivity of the transmission coefficients, anomalous
backscattering enhancement, and speckle pattern both in reflection and
transmission, reflectance and transmittance, and also in the conductance
and its universal fluctuations. It is shown that, in contrast to  volume
impurities, surface scattering in quasi-one-dimensional structures
(waveguides) gives rise to the coexistence of the ballistic, diffusive,
and localized regimes within the same sample.

\end{abstract}
\pacs{}%72.10.Fk, 73.23.-b, 42.25.Bs, 41.20.Jb
 }
\section{Introduction}

The statistical properties of disordered systems is a rich and long-standing 
problem which attracts many efforts both theoretical and experimental. In 
quantum solid state physics much attention is paid to statistics of 
eigenfunctions and eigenvalues of closed disordered systems. Disorder in 
solid state problems is usually represented by impurities which are randomly 
distributed over the whole sample. For this sort of mesoscopic samples with 
''bulk disorder'' the number of well established statistical characteristics 
is enormous (see, for example, Ref. \onlinecite{Efbook} and references
therein). The success of mesoscopics may be ascribed to the existence of
the non-linear $\sigma $-model put forward by Efetov \cite{Ef}. Although
it is a remarkable tool for studying mesoscopic effects, the $\sigma
$-model has, however, restricted validity. For example, the system size
must be much greater than the mean free path. Generalization to chaotic
ballistic systems (i.e. quantum billiards) has recently become a topic
of a great interest.  Progress in this direction has become possible due
to the recently proposed field theory for quantum ballistics
\cite{MKh}. By exploiting the new tool, the authors of
Refs. \onlinecite{tkh}  and \onlinecite{bmm} 
managed to treat different correlators in a clean system within an
extremely chaotic limit, when the typical relaxation time is of the
order of the flight time (diffusive boundary scattering). They found
that naive substitution of the mean free path for the system size into
correlators obtained for bulk disorder would give wrong results for the
ballistic case, and  that, in fact, systems with bulk and surface
disorder are not equivalent. 
 
In this paper we address the problem of a ballistic system that is
disordered in the sense that there are no bulk scatterers, and the only
place where scattering  
occurs is at a rough boundary. We choose to explore not the statistics
of the eigenvalues of completely closed systems (like resonators or
quantum dots), but the statistical properties of scattering and
reflecting amplitudes in bounded, but open in one direction, systems
(waveguiding structures). The key entities for this problem are the
transmission and reflection amplitudes. More specifically, we consider a
$N$-mode waveguide with the boundary corrugated within a finite interval
$(L),$ and study the statistical properties of the transmission through,
and reflection from, the disordered segment of the waveguide. This
problem arises naturally in the characterization of transport properties
related to, for instance, optical waveguides and fibers, remote sensing,
radio wave propagation, sonar, shallow water waves, and geophysical
probing \cite{sheng,frei}. On the other hand, it describes as well the
electronic transport in mesoscopic systems \cite{sheng95,meso}, being
especially relevant to conductance in nanowires 
\cite{taan,nimac,pedro,mole}.
 
The analogous problem with bulk scatterers was addressed by many authors
(see, for example, Refs. \onlinecite{sheng,sheng95,rev} and references
therein).
For a waveguide with bulk disorder all transmission coefficients $T_{mn}$
(sub-indexes $n$ and $m$ stand for number of the incident and transmitted
mode, respectively) behave in a similar way. Due to the strong inter-mode
mixing all information about the  $n$-dependence is washed out after a
few scattering events, which means that for $L$ of the order of the
scattering length $\ell$ (and larger) all modes cross over from the
ballistic regime to diffusion, and all channels become identical. As a
result, there exists only one characteristic length scale for all
transmittances, the so called localization length, that is believed to
be equal to $N\ell$. In the case $L\ll N\ell$, each $T_{mn}$ obeys
Rayleigh statistics; as the length $L$ increases, all 
channels (modes) undergo the same changes, and at $L\gg N\ell$ the
crossover to the log-normal distribution (typical for the localized
regime) takes place \cite{edrei,boer,nieu,kogan,vanl,stoy,brou}.
It might seem that the only distinction of the problem with a rough
surface from that with bulk disorder is that the scattering process
takes place in a reduced effective volume which should lead just to a
decrease of the mode mixing rate. Naive considerations would suggest
that if we introduced a new localization length (which obviously must be
much longer than that for the bulk scattering), all results well known
for the ``bulk'' problem should be valid for the ``surface'' case after
proper rescaling. However the situation is different and much more
complicated. 
 
The goal of the present paper is to study the statistical properties of
waves transmitted through and reflected from a waveguide with rough
boundaries. The length dependences of the reflection and transmission
amplitudes for each realization of the surface profile are numerically
obtained by solving a system of linear differential equations based on
the invariant embedding equations \cite{emb}. Then we calculate the
ensemble average and fluctuations of the reflection and transmission
coefficients, reflectance, transmittance, and conductance. It is shown
in this paper that the interplay between mode dispersion and surface
scattering gives rise to many new and unusual (at least from the point
of view of the intuition gained from studies of the volume scattering)
effects; one of such effects, the coexistence of different transport
regimes at a certain length scale, has been previously reported
\cite{prl98}. 

This paper is organized as follows. In Sec.~\ref{sec_the} the
theoretical formulation leading to the invariant embedding equations for
the matrices of the reflection and transmission amplitudes is
developed. The description of the numerical implementation of those
equations for the particular waveguide geometry chosen here is detailed
in Sec.~\ref{sec_num}. The results thus obtained for the average and
fluctuations of the reflection and transmission coefficients are
presented and discussed in Sec.~\ref{sec_rtmn}, whereas those for the
total reflection and transmission, and conductance, appear in
Sec.~\ref{sec_rtng}. The relevant conclusions derived from this work are
summarized in Sec.~\ref{sec_con}.

\section{Theory}

\label{sec_the}

We start from the wave equation
\begin{equation}
\left( \Delta +k^2\right) \Psi \left( {\bf R}\right) =0,
\label{eq_we}
\end{equation}
with the boundary conditions
\begin{mathletters}\begin{eqnarray}
\Psi\left({\bf R}={\bf R_s}\right)= & & 0, \text{ for } x<0
  \text{ and } x>L,\\  
= & & {\displaystyle -\mbox{\boldmath $\xi$}\left({\bf R}\right)\cdot
 \frac{\partial \Psi \left({\bf R} \right) }{\partial {\bf R}},} 
  \text{ for } 0\leq x\leq L,
\end{eqnarray}\label{eq_bc}\end{mathletters}
given on the unperturbed waveguide surface ${\bf R}={\bf R_s}$, which is
translationally invariant along the $x$-axis 
(${\bf R}=\left( x,{\bf r}\right)$). The boundary condition (\ref{eq_bc})
corresponds either to a waveguide surface with a random admittance 
{\boldmath $\xi$}, or to the Dirichlet boundary condition on a slightly
perturbed  waveguide surface, {\boldmath $\xi$} denoting the random
perturbation. In the latter case the boundary condition (\ref{eq_bc}) is
an approximate one, containing the first two terms in the expansion of
the exact (Dirichlet) boundary condition about the unperturbed surface.

Outside the region $0\leq x\leq L$ the solutions of the scattering
problem under consideration have the form:
\begin{mathletters}\begin{eqnarray}
\Psi _n\left( x,{\bf r}\right) = & {\displaystyle \sum_m
 \frac 1{\sqrt{k_m}}\chi_m\left( {\bf r}\right) e^{-ik_mx}t_{mn},} & 
 \quad x<0, \\ 
\Psi _n\left( x,{\bf r}\right) =& {\displaystyle \frac 1{\sqrt{k_n}}\chi
_n\left( {\bf r}\right) e^{-ik_nx}\hspace*{1.5cm}} & \nonumber \\ &
{\displaystyle +\sum_m\frac 1{\sqrt{k_m}}\chi_m\left( {\bf r}\right)
  e^{ik_mx}r_{mn}, } & \quad x>L. 
\end{eqnarray}\label{eq_sc}\end{mathletters}
The indexes ``$m,n$'' correspond to the outgoing and incoming modes,
respectively, and $\chi _n\left( {\bf r}\right)$ are the
eigenfunctions of the transverse wave equation
\begin{equation}
\left( \frac{\partial ^2}{\partial {\bf r}^2}+\kappa _n^2\right)
\chi _n\left( {\bf r}\right) =0,  \label{eq_twe}
\end{equation}
with $k_n=\sqrt{k^2-\kappa _n^2}$. 

By assuming that {\boldmath $\xi$}$=\xi{\bf n}$, the boundary condition
(\ref{eq_bc}) can be also rewritten as
\begin{equation}
\Psi \left( {\bf R_s}\right) +\xi \left( {\bf R_s} \right) \Phi \left(
 {\bf R_s}\right) =0,  
\label{eq_bc1} 
\end{equation}
where the normal derivative 
\begin{equation}
\Phi\left( {\bf R_s}\right) \equiv{\bf n}\left( {\bf R_s}\right)\cdot
 \left[\frac{\partial\Psi \left( {\bf R}\right) }{\partial {\bf R}}
 \right]_{{\bf R}={\bf R_s}}\equiv \frac{\partial \Psi }{\partial n}
\label{eq_nd}
\end{equation}
is introduced, with ${\bf n}\left( {\bf R_s}\right) $ being the normal
to the unperturbed surface ${\bf R}={\bf R_s}.$

Let us employ Green's theorem in the form:
\begin{eqnarray}
\Psi \left( {\bf R}\right) = & {\displaystyle \int d^3R^{\prime }
 \frac \partial {\partial {\bf R^{\prime }}}\cdot\left[ \Psi \left( {\bf
R^{\prime }}\right) \frac{\partial G_0\left( {\bf R^{\prime }},
{\bf R}\right) }{\partial {\bf R^{\prime }}}\right.} &\nonumber \\
 &{\displaystyle\left. -\frac{\partial \Psi \left( {\bf R^{\prime
}}\right)}{\partial{\bf R^{\prime }}}G_0\left( {\bf R^{\prime }},{\bf
R}\right) \right] , }
\label{eq_gt}
\end{eqnarray}
where the integral is taken over any region containing the point 
${\bf R}$ and located inside the waveguide. The most convenient
integration region is the unperturbed waveguide surface cut by two
planes normal to the axis of the waveguide (they are included too). The 
left (right) cross-section must be placed to the left (right) of
the point ${\bf R}=\left( x,{\bf r}\right)$. The integral in Eq.
(\ref{eq_gt}) can  then be expressed as the sum of the integrals over
the cross-sections and the unperturbed waveguide surface with constraint
$0\leq x^{\prime }\leq L$:
\begin{equation}
\Psi \left( {\bf R}\right) =\int d{\bf S^{\prime }}\cdot
\Psi \left( {\bf R_s^{\prime }}\right) \left[ \frac{\partial
G_0\left( {\bf R^{\prime }},{\bf R}\right) }{\partial
{\bf R^{\prime }}}\right] _{{\bf R^{\prime }}={\bf R_s^{\prime }%
}}+I_{cross}.  \label{eq_gti}
\end{equation}
The integral $I_{cross}$ over the cross-sections may be easily calculated
from Eqs. (\ref{eq_sc}) and the Green's function
\begin{equation}
G_0\left( {\bf R},{\bf R^{\prime }}\right)
=\sum_{m=1}^N\frac{\chi _m\left( {\bf r}\right) \chi _m\left(
{\bf r^{\prime }}\right) }{2ik_m}e^{ik_m\left| x-x^{\prime
}\right| }.  \label{eq_gf}
\end{equation}
Then  Eq. (\ref{eq_gti}) takes the form:
\begin{eqnarray}
\Psi _n\left( {\bf R}\right) =& &\frac 1{\sqrt{k_n}}\chi _n\left(
{\bf r}\right) e^{-ik_nx} \nonumber \\ & &
 +\int d{\bf S^{\prime }}\cdot\Psi_n\left( {\bf R_s^{\prime }}\right)
\left[ \frac{\partial G_0\left( {\bf R^{\prime }},{\bf R}\right)
  }{\partial {\bf R^{\prime }}}\right]
  _{{\bf R^{\prime }}={\bf R_s^{\prime }}}  .
\label{eq_gti1}
\end{eqnarray}
The (oriented) surface element $d{\bf S}$ can be written as $d%
{\bf S}={\bf n}dS$, $dS=dx\;ds$. The explicit form
of the differential $ds$ depends on the geometry under consideration 
($ds=rd\phi$ for circular cross-sections, $dydz$ for rectangular
cross-sections, etc.). Then we can rewrite Eq. (\ref{eq_gti1}) as:
\begin{eqnarray}
\Psi _n\left( {\bf R}\right) = && \frac 1{\sqrt{k_n}}\chi _n\left(
{\bf r}\right) e^{-ik_nx} \nonumber \\ & &
 +\int\limits_0^Ldx^{\prime }\oint
 ds^{\prime }\Psi _n\left( x^{\prime },{\bf r_s^{\prime }}\right)
\frac{\partial G_0\left( x^{\prime },x;{\bf r_s^{\prime }},%
{\bf r}\right) }{\partial n^{\prime }}  \label{eq_gti2}
\end{eqnarray}
Two conclusions can be derived from Eq. (\ref{eq_gti2}). First, the
matrix of the reflection coefficients can be written [after substituting
the explicit expression for $G_0$ from Eq. (\ref{eq_gf})] in the form:
\begin{equation}
r_{mn}=\frac 1{2i}\int\limits_0^Ldx\oint ds\frac
1{\sqrt{k_m}}\phi _m\left( {\bf r}\right) e^{-ik_mx
}\Psi _n\left( x,{\bf r_s}\right),  \label{eq_rc}
\end{equation}
or [by the use of Eq. (\ref{eq_bc1})] as:
\begin{equation}
r_{mn}=-\frac 1{2i}\int\limits_0^Ldx\oint ds\;\phi _m\left( x,s\right)
\xi\left( x,s\right) \Phi _n\left( x,s\right),  \label{eq_rc1}
\end{equation}
where
\[
\phi _n\left( x,s\right) =\frac 1{\sqrt{k_n}}{\bf n}({\bf r_s})\cdot
\left[ \frac{\partial \chi
_n\left( {\bf r}\right) }{\partial {\bf r}}\right] _{{\bf r}={\bf
r_s}}e^{-ik_nx} .
\]
The second result we derive from Eq. (\ref{eq_gti2}) [by differentiating,
setting ${\bf R}$ on the surface, and substituting the boundary
condition (\ref{eq_bc1})] is a closed equation for
$\Phi_n\left( x,s\right)$:
\begin{eqnarray}
&&\Phi _n\left( x,s\right) =  \phi _n\left( x,s\right) \nonumber \\ &&
-\int\limits_0^Ldx^{\prime }\oint ds^{\prime }\xi \left( x^{\prime
},s^{\prime }\right) \Phi _n\left( x^{\prime },s^{\prime }\right)
G_0^{^{\prime \prime }}\left( x^{\prime },x;s^{\prime },s\right). 
\label{eq_ce}
\end{eqnarray}
Here $G_0''$ is the mixed normal derivative
\[
G_0^{^{\prime \prime }}\left( x^{\prime },x;s^{\prime },s\right)
 =\frac{\partial ^2G_0}{\partial n\partial n^{\prime }}=\frac 1{2i}\sum
 \frac 1{k_n}\frac{\partial \chi _n}{\partial n}\frac{\partial \chi
 _n}{\partial n^{\prime }}e^{ik_n\left| x-x^{\prime }\right| }.
\]
Differentiation of Eq. (\ref{eq_ce})  yields
\begin{eqnarray}
 &&\frac{\partial\Phi _n\left( x,s\right)}{\partial L}=\sum_m\phi_m
 \left( x,s\right) a_{mn} \nonumber \\ &&
 -\int\limits_0^Ldx^{\prime }\oint ds^{\prime}
 \xi\left( x^{\prime},s^{\prime}\right) \frac{\partial \Phi_n\left( 
 x^{\prime},s^{\prime}\right) }{\partial L}G_0^{^{\prime\prime}}\left(
 x^{\prime},x;s^{\prime},s\right) ,  \label{eq_ed} \\
&& a_{mn} =-\frac 1{2i}\oint ds^{\prime }\phi _m^{*}\left( L,s\right)\xi
 \left( L,s\right) \Phi _n\left( L,s\right).
\end{eqnarray}
By comparing Eq. (\ref{eq_ed}) with Eq. (\ref{eq_ce}) we obtain the
relation between the derivative $\partial \Phi_n(x,s)/\partial L$ and
the functions $\{\Phi_m(x,s)\}$: 
\begin{equation}
\frac{\partial \Phi _n\left( x,s\right) }{\partial L}=\sum_m\Phi_m
 \left( x,s\right) a_{mn} . \label{eq_i}
\end{equation}
With the aid of the latter equation, we now differentiate the matrix of
reflection coefficients Eq. (\ref{eq_rc1}):
\begin{eqnarray}
&&\frac{dr_{mn}}{dL} =-\frac 1{2i}\oint ds\;\phi _m\left( L,s\right) \xi
\left( L,s\right) \Phi _n\left( L,s\right)    \nonumber\\
&&\ -\frac 1{2i}\int\limits_0^Ldx\oint ds\;\phi _m\left( x,s\right) \xi
\left( x,s\right) \sum_m\Phi _m\left( x,s\right) a_{mn} . \label{eq_dri} 
\end{eqnarray}
By substituting the explicit expressions for $\Phi _m\left( x,s\right) $ 
from Eqs. (\ref{eq_sc}b) and (\ref{eq_nd}), and
collecting all terms, we arrive at
\begin{mathletters}
\label{eq_iee}
\begin{equation}
\frac{d\widehat{r}}{dL}=\frac i2\left( e^{-i\widehat{k}L}+\widehat{r}
 e^{i\widehat{k}L}\right) \widehat{v}\left( e^{-i\widehat{k}L}+
 e^{i\widehat{k}L}\widehat{r}\right) . \label{eq_ier}
\end{equation}
Here $\widehat{k}=diag\left( k_n\right) $ and
\begin{eqnarray}
v_{mn}=\oint ds\,\phi _m\left( s\right) \xi \left( L,s\right) \phi
_n\left(s\right) , \nonumber \\
\phi _n\left( s\right) =\frac 1{\sqrt{k_n}}{\bf n}({\bf r_s})\cdot
\left[ \frac{\partial \chi _n\left( {\bf r}\right)
}{\partial {\bf r}}\right] _{{\bf r}={\bf r_s}}. \nonumber
\end{eqnarray}

Analogous algebra leads to the equation for the matrix of transmission
coefficients:
\begin{equation}
\frac{d\widehat{t}}{dL}=\frac i2\widehat{t}e^{i\widehat{k}L}\widehat{v}
\left( e^{-i\widehat{k}L}+e^{i\widehat{k}L}\widehat{r}\right) .
\label{eq_iet}
\end{equation}
\end{mathletters}

From the reflection and transmission amplitudes, we define the
reflection and transmission intensities, respectively,
\begin{eqnarray}
R_{mn}=\left| r_{mn}\right| ^2, T_{mn}=\left| t_{mn}\right| ^2,
\end{eqnarray}
which yield the intensity coupled into the $mth$ outgoing channel in
reflection and transmission, respectively, for a given $nth$ incoming
channel. The reflectance and transmittance for the $nth$ incident mode
are
\begin{eqnarray}
R_n=\sum\limits_m R_{nm},\hspace*{.3in} T_n=\sum\limits_m T_{nm}.
\end{eqnarray}
Finally, the total transmitted intensity in the case that all incoming
channels are incoherently populated, which is equivalent to the
dimensionless conductance for electrons, is
\begin{eqnarray}
g=\sum\limits_n T_n.
\end{eqnarray}

\section{Numerical Calculations}
\label{sec_num}

\begin{figure}
\epsfbox{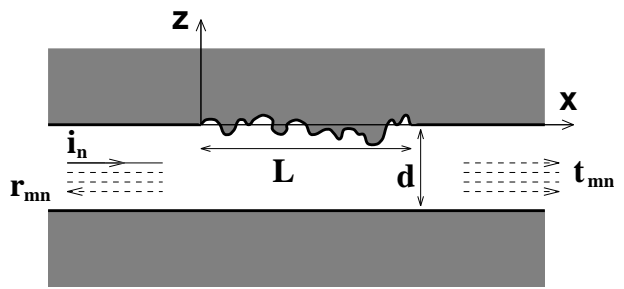}
\caption{Illustration of the waveguide geometry.}
\label{fig_wg}
\end{figure}

For the numerical simulations we choose the simplest geometry (see
Fig.~\ref{fig_wg}): two parallel planes $z=0$ and $z=d$ with 1D
deviations $\xi =\xi \left( x\right) $ on one plane $\left( z=0\right)$
only, where $\xi$ is  a 1D stochastic process. Thus the transverse
eigenfunctions acquire the form 
\begin{equation}
\chi_n\left( z\right) =\sqrt{\frac 2d}\sin\left( \kappa _nz\right) ,\;
  \kappa _n=\frac{\pi n}d,\; k_n=\sqrt{\left( \frac \omega c\right)^2
  -\kappa _n^2},
\end{equation}
and the impurity matrix becomes 
\begin{equation}
v_{mn}\left( L\right) =\frac 2d\frac{\kappa _n\kappa _m}{(k_nk_m)^{1/2}}
\xi\left( L\right) . \label{eq_im}
\end{equation}

The $2N\times 2N$ system of linear differential equations (\ref{eq_iee})
is solved numerically  by means of the 6th-order Runge-Kutta method. For
each realization $\xi(x)$ (of length $L_{max}$) of an ensemble of
randomly rough surface profiles, the matrices of reflection and
transmission amplitudes are calculated as  functions of the length
$L$. These realizations obey Gaussian statistics (with $\delta$ the rms
height) with zero mean and a Gaussian correlation function 
\begin{equation}
W(|x-x'|)=\delta^{-2}\langle\xi(x)\xi(x')\rangle=
        \exp\left(-a^{-2}(x-x')^2\right), 
\label{eq_wx}
\end{equation}
where $a$ is the transverse correlation length. The corresponding
surface power spectrum is thus given by:
\begin{equation}
g(Q)=\pi^{\frac{1}{2}}a \exp\left(-(Qa)^2/4\right).
\label{eq_ps}
\end{equation}
The ensemble of surface realizations are numerically generated as
described in Ref. \onlinecite{jasg}. By averaging over $N_p$
such realizations, the mean values $\langle A\rangle$ and fluctuations
$\delta A=(\langle A^2\rangle -\langle A\rangle^2)^{1/2}$  of the
relevant physical quantities are obtained. Hereafter we consider, unless
otherwise stated, a waveguide of thickness $d=2.25\lambda$ supporting
$N=4$ guided modes. 

\section{Reflection and transmission coefficients}
\label{sec_rtmn}

\begin{figure}
\epsfxsize=2.5in \epsfbox{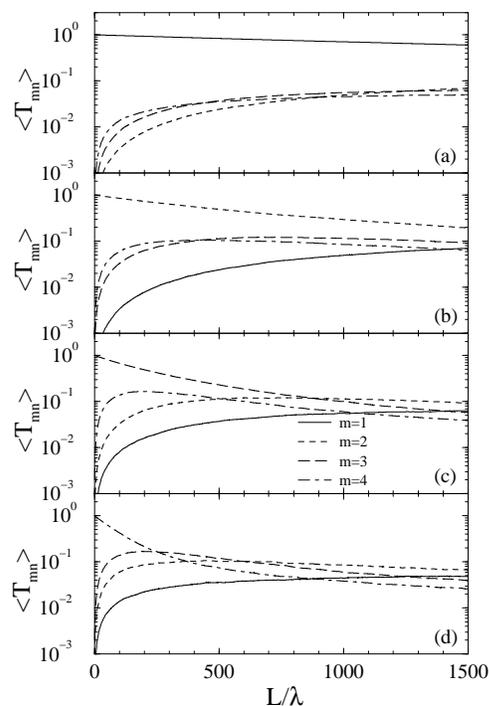}
\caption{Mean transmission intensities $\langle T_{mn}\rangle$ as
functions of length $L$ in semi-log scale for a waveguide of
width $d/\lambda=2.25$, supporting 4 modes, with disorder parameters
$a/\lambda=0.2$ and $\delta/\lambda=0.03$: (a) incident mode $n=1$; (b)
$n=2$; (c) $n=3$; (d) $n=4$. Averaged over $N_p=4000$ realizations.}\
\label{fig_tmn}
\end{figure}

\begin{figure}
\epsfxsize=2.5in \epsfbox{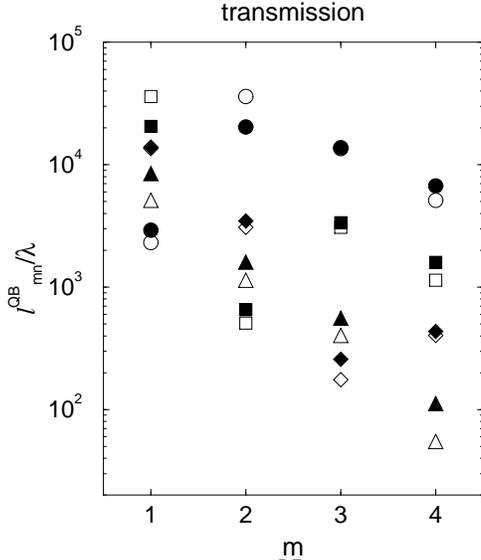}
\caption{QB lengths $\ell^{QB}_{mn}$ (in wavelength units) in
transmission versus outgoing channel $m$ for the waveguide used in
Fig. \protect{\ref{fig_tmn}}: Circles: $n=1$; Squares: $n=2$; Diamonds:
$n=3$; Triangles: $n=4$. Filled (respectively, open) symbols denote the
numerical simulation (respectively, perturbation theory) results.}
\label{fig_lqbmnt}
\end{figure}

In Fig.~\ref{fig_tmn} the $\langle T_{mn}\rangle$ are shown (in a
semi-log scale) for $L\leq L_{max}=1500\lambda$. Averaging was carried
out over the results obtained for $N_p=4000$ realizations of the surface
profile, whose roughness parameters are $a=0.2\lambda$ and
$\delta=0.03\lambda$. The asymmetry in the behavior of the different
outgoing channels $m$ is evident from this plot. The intensity of the
incoming mode $\langle T_{nn}\rangle$ 
decreases with length for all $n$, this decrease being steeper the
larger the transverse momentum $\kappa_n$, namely, the larger $n$.
The transmission into other non-diagonal channels $m\not =n$ also
depends strongly on the mode $m$. In the beginning of the waveguide,
this non-diagonal transmission slightly increases from zero, being
stronger into higher modes $m$ ($\langle T_{m+1,n}\rangle>\langle
T_{mn}\rangle$, with $m+1,m\not =n$). In this situation only  {\it
single} scattering is important, and we refer to this regime as
quasi-ballistic (QB). In accordance with the results of perturbation
theory (PT) the intensity of mode $m$  is proportional to the cross
section for roughness-induced scattering from mode $n$ into mode $m$,
and to the length $L$, as follows: 
\begin{equation}
\langle T_{mn}\rangle =\frac{2\delta ^{2}\kappa _{n}\kappa
_{m}}{d^{2}(k_{n}k_{m})^{1/2}}g(\left| k_{n}-k_{m}\right| )L\equiv 
\frac{L}{\ell^{QB}_{mn}}.
\label{eq_tmnqb}
\end{equation}
For the diagonal transmission, PT predicts
\begin{eqnarray}
\langle T_{nn}\rangle = && 1-L\frac{2\delta ^{2}}{d^2}\sum_{m=1}^{N}
\frac{\kappa _{n}\kappa_{m}}{(k_{n}k_{m})^{1/2}}
\left[g(\left|k_{n}-k_{m}\right| )\right. \nonumber \\ && 
 \left.+g(\left|k_{n}+k_{m}\right| )\right]\equiv
  1-\frac{L}{\ell^{QB}_{nn}}. 
\label{eq_tnnqb}
\end{eqnarray}
In Fig.~\ref{fig_lqbmnt}  the corresponding QB lengths $\ell^{QB}_{mn}$
from the preceding PT expressions have been plotted  along with those
obtained by fitting the numerical results shown in Fig.~\ref{fig_tmn}
to the expected linear functions, showing good agreement.

The origin of such asymmetries lies in the surface-type disorder that
randomizes the wave propagation through the waveguide. If we look at the
impurity matrix (\ref{eq_im}), which determines the scattering strength
in Eqs. (\ref{eq_iee}), it is obvious that there are large
quantitative differences in $v_{mn}$ for distinct values of $m$ and $n$.
As a matter of fact, this matrix can be rewritten as:
\begin{equation}
 v_{mn}= \frac{8d^2}{L^2}\frac{\xi(L)}{d}(k_{n}k_{m})^{1/2}\, M_m M_n ,
\label{eq_vmn}
\end{equation}
where $M_n$ is 
\begin{equation}
         M_n = \frac{L\kappa_n}{2dk_n}.
\label{eq_mn}
\end{equation}
Through a simple geometrical argument, as long as $\delta\ll d$, $M_m$
can be considered the number of times that mode $m$ hits (interacts
with) the rough wall on its way along the waveguide \cite{prl98}. In the
case that $d/\lambda=2.25$, it turns out that, for instance,
$M_4\approx 8.5M_1$. This factor affects the impurity matrix not only
for the outgoing mode through $M_m$, but also for the incoming mode 
through $M_n$. This gives a physically intuitive explanation of the
results shown in Figs.~\ref{fig_tmn}, and of all other processes
that will be shown below.

For larger $L$ {\it multiple} scattering becomes relevant. This actually
means that not only the scattering that brings energy to mode $m$ from
$n$ should be taken into account, but also the leakage from $m$ into
other modes, as well as all interchanges between $i$ and $j$ for all
$i,j$. As a result, the energy spreads over all modes:  diffusion (D) 
in the space of mode-numbers takes place \cite{bass}. In fact, it is
seen in Fig.~\ref{fig_tmn} that all outgoing channels tend to yield
comparable transmission intensities within the length of the plot,
except for $n=1$. 

\begin{figure}
\epsfxsize=2.5in \epsfbox{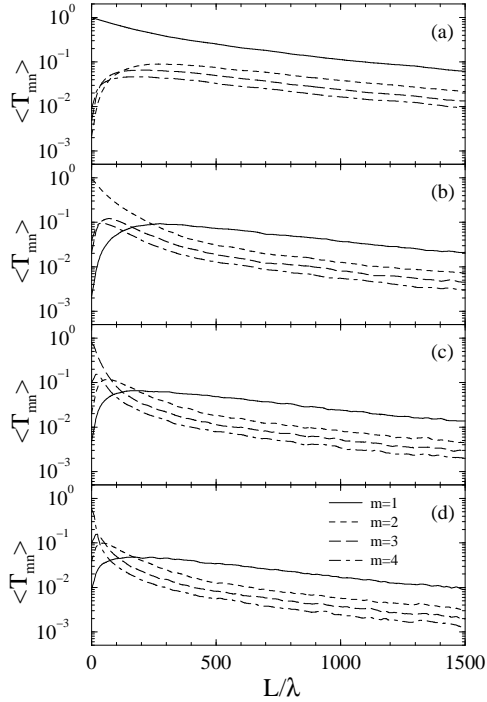}
\caption{Same as in Fig. \protect{\ref{fig_tmn}} but for
$\delta/\lambda=0.1$.}
\label{fig_tmn1}
\end{figure}

Furthermore, after a  long propagation distance through the waveguide so
that mode conversion has sufficiently populated all outgoing channels, we
observe that $T_{1n}>T_{2n}>T_{3n}>T_{4n}$. This waveguide length is not
reached within the length scale covered in Fig.~\ref{fig_tmn}.
Alternatively, we have increased the surface roughness to $\delta=0.1
\lambda$, so that this result can be observed: the corresponding
transmission coefficients, presented in Fig.~\ref{fig_tmn1}, indeed
confirm such behavior for a waveguide length $L\geq 300\lambda$. Thus, 
the higher modes appear to be more strongly scattered. This is also
manifested in the overall behavior of the four outgoing channels
depending on the incident channel. Figure~\ref{fig_tmn1} reveals that,
beyond the waveguide length given above, the transmission curves appear
to be qualitatively similar for all incoming modes $n$, but shifted
downward as $n$ is increased. The behavior of those transmission curves,
following parallel exponential decays, is a signature of the onset of
localization (L) due to the coherent interference of multiply scattered
waves. (On the other hand, it is interesting to note that, if we zoom in
Fig.~\ref{fig_tmn1} for $0<L<100\lambda$, the transmission curves are
qualitatively similar to those of Fig.~\ref{fig_tmn}.)

\begin{figure}
\epsfxsize=2.5in \epsfbox{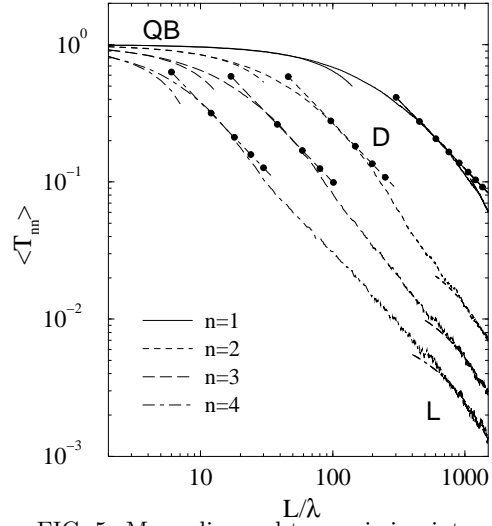}
\caption{Mean diagonal transmission intensities 
$\langle T_{nn}\rangle$  as in Fig. \protect{\ref{fig_tmn1}} in 
a log-log plot. Fits to linear (QB), $L^{-1}$  (D, with dots), and
exponential (L) decays are shown.}
\label{fig_tnn}
\end{figure}

Thus we have seen in Figs.~\ref{fig_tmn} and~\ref{fig_tmn1}  that the
dependence of the impurity matrix (\ref{eq_vmn}) on mode dispersion has
significant quantitative consequences, and  also strong qualitative
consequences for the properties of wave propagation through 
surface-disordered waveguides. As has been demonstrated in Ref.
\onlinecite{prl98}, it can give rise to an entangling of transport
behaviors within the same waveguide length. In Fig.~\ref{fig_tnn}, the
diagonal transmission coefficients $T_{nn}$ from Fig.~\ref{fig_tmn1} are
shown in a log plot. The results have been fitted, where possible, to
the well-known behaviors: QB as in Eq. (\ref{eq_tnnqb}); inverse power
law expected for D
\begin{equation}
      \langle T_{nn}\rangle\approx \frac{l^D_{nn}}{L};
\label{eq_d}
\end{equation}
and exponential decay associated with L.
It is seen in Fig.~\ref{fig_tnn} that, within the interval $10<L/
\lambda<70$, QB transport of the (11) channel coexists with D for the
(33) and (44) channels; also, D of the (11) mode coexists with L of the
(44) channel for 
$L/\lambda\approx 10^3$. This confirms the coexistence of QB transport,
D, and L predicted in Ref.~\onlinecite{prl98} for 8-mode waveguides;
nonetheless, in this 4-mode waveguide the coexistence of all three
regimes within the same length region is not observed due to the
limitation in mode dispersion differences associated with the lower
number of available modes. [On the other hand, it should be noted that
our results, not shown here, reveal such coexistence (QB-D or D-L)
phenomena associated with surface-type disorder in the case of narrower
waveguides supporting only 3 or even 2 guided modes.] Interestingly, the
impossibility of defining the D regime consistently for all outgoing
modes at the same length scale makes irrelevant any comparison with
theories such as the macroscopic approach provided by random matrix
theory (RMT)\cite{rev,rmt}, which predicts $\langle T_{mn}\rangle=
\langle g\rangle/N^2$ for all $m,n$.

In addition, Figure~\ref{fig_tnn} permits us to observe the crossover
between different regimes for each mode separately. For $L<\ell^{QB}_{44}$,
all 4 modes propagate almost ballistically. The transition from
ballistic transport to D can be observed for all modes at the
distinct waveguide lengths defined by the corresponding $\ell^D_{nn}$ (see
Fig.~\ref{fig_lnn} below). Note that even though QB and D regimes extend
over different $L$ regions, in both cases the regions are well defined
by the magnitude of the transmission coefficient: $\langle T_{nn}\rangle
\approx 1$ for QB and $\langle T_{nn}\rangle\approx 10^{-1}$ for D.
This seems to indicate that, from the value of the average transmission
coefficient, the qualitative transport behavior can be roughly known, in
agreement with Ref. \onlinecite{mnvprl}, although there exist remarkable
differences concerning the length dependence and the entangling 
of regimes. Finally, coherent interference leads to L. In
Figure~\ref{fig_tnn} all modes [mode (11) barely] reach the L regime
within the maximum length of the waveguide $L_{max}$. It should be
remarked that, whereas the exponential decay rate is similar for all
$n$, the real onset of localization takes place at slightly different
lengths: the lower $n$ is, the longer the waveguide must be to observe
L.

The dependence of $\ell^{QB}_{nn}$, $\ell^D_{nn}$, and $\ell^L_{nn}$ on
surface roughness  is shown in Fig.~\ref{fig_lnn}. In this respect, with
the aim of correctly defining $\ell^L_{nn}$, the average of the
logarithm of the transmission has been used \cite{frei}: 
\begin{equation}
      \langle\ln T_{nn}\rangle\approx -\frac{L}{\ell^L_{nn}}.
\label{eq_l}
\end{equation}
The predicted $\delta^{-2}$ behavior is seen in Fig.~\ref{fig_lnn}(a)
for the QB decay lengths, showing reasonable agreement with the PT
results [cf. Eq. (\ref{eq_tnnqb})]. Similar behavior is observed in 
Fig.~\ref{fig_lnn}(b) for $\ell^D_{nn}$ and $\ell^L_{nn}$. It is
interesting to note that $\ell^D_{nn}$ is different for each $n$,
whereas $\ell^L_{nn}$ coincides for all $n$. Thus the well known
relationship $\ell^L=N\ell^D$ is meaningless in this context. (Although
$\ell^L=N\ell^D_{11}$ seems to hold instead; in fact, it has been shown
that if $\ell^D$ is defined through the resistance,  $\ell^L=N\ell^D$
does hold \cite{mnvapl}.)

\begin{figure}
\epsfxsize=2.5in \epsfbox{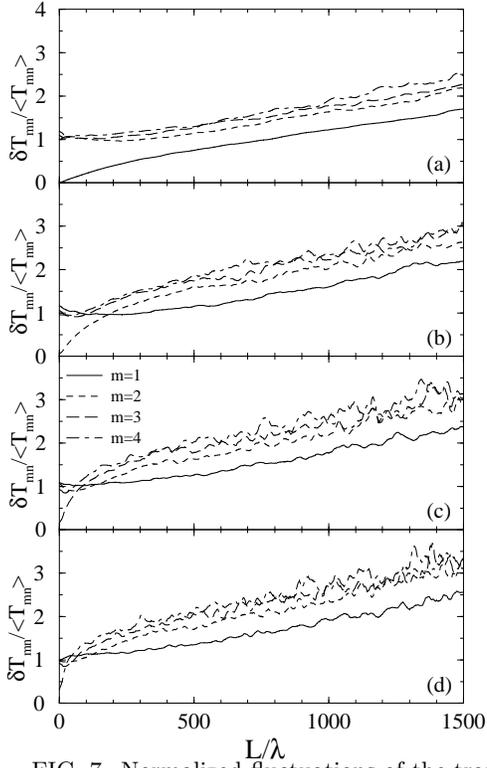}
\caption{Normalized fluctuations of the transmission intensities as 
functions of the length $L$, for the same parameters as in
Fig.~\protect{\ref{fig_tmn1}}.}
\label{fig_dtmn}
\end{figure}

The normalized fluctuations $\delta T_{mn}/\langle T_{mn}\rangle$ are
shown in Fig.~\ref{fig_dtmn}. It is evident that there are differences
among the fluctuations for every channel, in agreement with the behavior
of the mean values shown in Fig.~\ref{fig_tmn1}; this corroborates the
qualitative argument given above in connection with the asymmetry in the
mode scattering rates. Note that at the beginning of the waveguide, mode
conversion into $m\neq n$ leads to a variance of unity for the
corresponding off-diagonal fluctuations, whereas the diagonal $m=n$
ballistic transport is revealed through the result that 
$\langle\delta T_{nn}\rangle\approx 0$. 
Furthermore, these diagonal fluctuations undergo the crossovers between
QB, D, and L regimes as discussed above in light of the mean values (see
Fig.~\ref{fig_tnn}). The $(nn)$-fluctuations exhibit an increase  from 0
towards 1 as the transport gradually changes from QB to D. The well
known speckle pattern fluctuations $\delta T_{nn}/\langle T_{nn}\rangle
\approx 1$ for all $(mn)$-channels build up in the D regime, steadily
increasing above 1 as the mode becomes localized. Therefore, the
phenomenon of the QB-D and D-L coexistence  can be recognized by
comparing the diagonal fluctuations with each other, corroborating the
argument given above in light of the results for the mean values in 
Fig.~\ref{fig_tnn}. For a sufficiently long waveguide, it can be seen in
Fig.~\ref{fig_dtmn} that the normalized fluctuations tend to be larger
the higher the outgoing mode $m$ is. A linear increase for all channels
is observed \cite{taan}. For a given incoming mode $n$, the rate of
increase is the same for all outgoing channels $m$; nonetheless, the
fluctuations appear to be larger the higher $m$ is (within the noise
accuracy). Analogously, the rate of increase  is faster the higher $n$
is. These considerations corroborate the  arguments discussed above on
the mode selectivity of the scattering strength in connection with the
transmission intensities in Fig.~\ref{fig_tmn1}.

\begin{figure}
\epsfxsize=2.5in \epsfbox{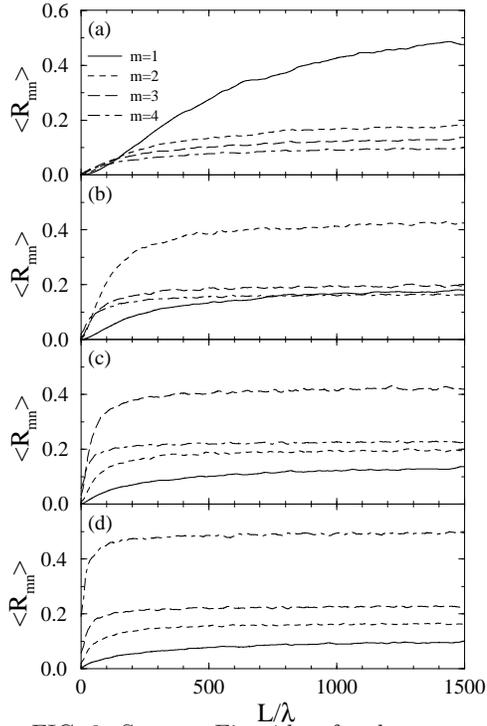}
\caption{Same as Fig. \protect{\ref{fig_tmn1}} but for
the mean reflection intensities $\langle R_{mn}\rangle$.}
\label{fig_rmn}
\end{figure}

\begin{figure}
\epsfxsize=2.5in \epsfbox{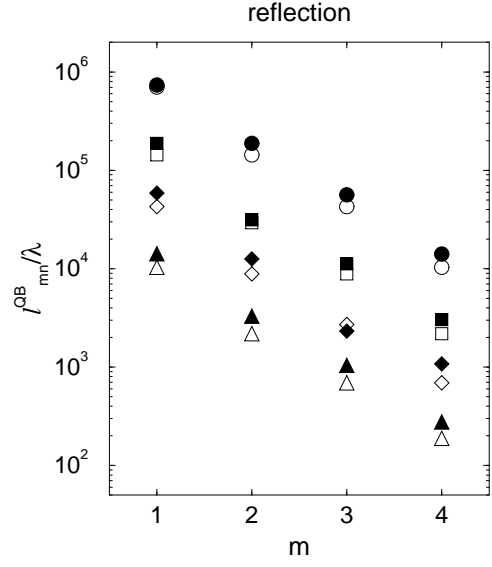}
\caption{Same as Fig. \protect{\ref{fig_lqbmnt}} but in
reflection.}
\label{fig_lqbmnr}
\end{figure}

Let us now turn to the study of the reflection coefficients $\langle
R_{mn}\rangle$. These are presented in Fig.~\ref{fig_rmn} for the same
waveguide considered in Fig.~\ref{fig_tmn1}. The peculiar scattering
properties induced by surface disorder manifest themselves in an
intricate manner in the reflection channels also. For sufficiently short
waveguide lengths, we expect that the reflection coefficients should
increase linearly as predicted by PT, through the expressions:
\begin{equation}
\langle R_{mn}\rangle =\frac{2\delta ^{2}\kappa _{n}\kappa
_{m}}{d^{2}(k_{n}k_{m})^{1/2}}g(\left| k_{n}+k_{m}\right| )L\equiv 
\frac{L}{\ell^{QB}_{mn}} .
\label{eq_rmnqb}
\end{equation}
These PT QB decay lengths in reflection and those obtained from the
numerical results  are shown in Fig.~\ref{fig_lqbmnr}. The agreement is
even better than in transmission, and the strong mode differences are
indeed confirmed. Beyond the QB regime for each incoming mode
$n$, the diagonal $\langle R_{nn}\rangle$ is enhanced as the waveguide
length increases, whereas the remaining off-diagonal reflection
coefficients exhibit differences with the following tendency: the higher
is the mode $m$, the larger is  the  reflection coefficient. Enhanced
backscattering appears in the strong diffusive (or weak localization)
regime as a result of the constructive interference of multiply
scattered paths; nonetheless, the enhancement factor differs from 2
(predicted by simple arguments, provided that the single scattering
contribution is absent or negligible). This comes as no surprise
inasmuch as each channel may behave differently, as revealed in the
transmission coefficients (see Fig.~\ref{fig_tmn1}) through the
entangling of transport regimes. Interestingly enough, this anomalous
reflection can result in an anomalous enhancement factor \cite{mnvprlr},
larger than 2 [see Fig.~\ref{fig_rmn}(a): although the background cannot
be unambiguously defined, $\langle R_{11}\rangle \approx 2.3\langle
R_{m1}\rangle$ for any $m\neq 1$]. Therefore the reflection coefficients
fail to satisfy \cite{rev,rmt} $\langle R_{mn}
\rangle=(1+\delta_{mn}) N^{-1}(1+N)^{-1}(N-\langle g\rangle)$.
Figure~\ref{fig_rmn} also seems to indicate that the onset of L does not
introduce significant changes in the reflection coefficients, in
agreement with Ref. \onlinecite{mnvprlr}.

\begin{figure}
\epsfxsize=2.5in \epsfbox{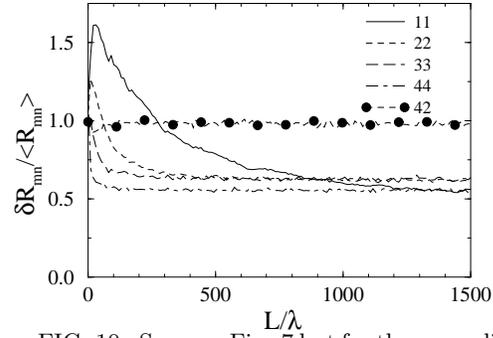}
\caption{Same as Fig. \protect{\ref{fig_dtmn}} but for the
normalized fluctuations of the reflection intensities, including only
the backscattered channels and the (42) off-diagonal channel.}
\label{fig_drmn}
\end{figure}

It is interesting to analyze the normalized fluctuations of the reflection
coefficients (see Fig.~\ref{fig_drmn}). Leaving aside the transient strong
fluctuations for very short length scales (associated with the fact that
the corresponding reflection coefficients are small), the diagonal $m=n$
normalized fluctuations diminish with increasing length, this decrease
being steeper the higher $n$ is. Then they stabilize about the value
$\delta R_{nn}/\langle R_{nn}\rangle=0.5$ as the D regime is reached, 
and remain constant when entering into the L regime. The off-diagonal
normalized fluctuations [only the (42) channel is shown in
Fig.~\ref{fig_drmn}), since all the rest are similar], on the other
hand, remain about the variance of unity linked to the speckle pattern
fluctuations in reflection. Therefore weak localization halves speckle
pattern fluctuations in backscattering.

\section{Reflectance, transmittance, and conductance}
\label{sec_rtng}

We have thus seen that wave propagation along a
4-mode surface-disordered waveguide, due to the surface-type disorder,
unlike for volume disorder, displays anomalous properties in the
transmission and reflection coefficients as a consequence of
the mixture of QB, D, and L regimes for different waveguide channels.
Bearing in mind these properties, we now proceed to calculate the
total transmission $T_n$, reflection $R_n$, and dimensionless
conductance $g$.

\begin{figure}
\epsfxsize=2.5in \epsfbox{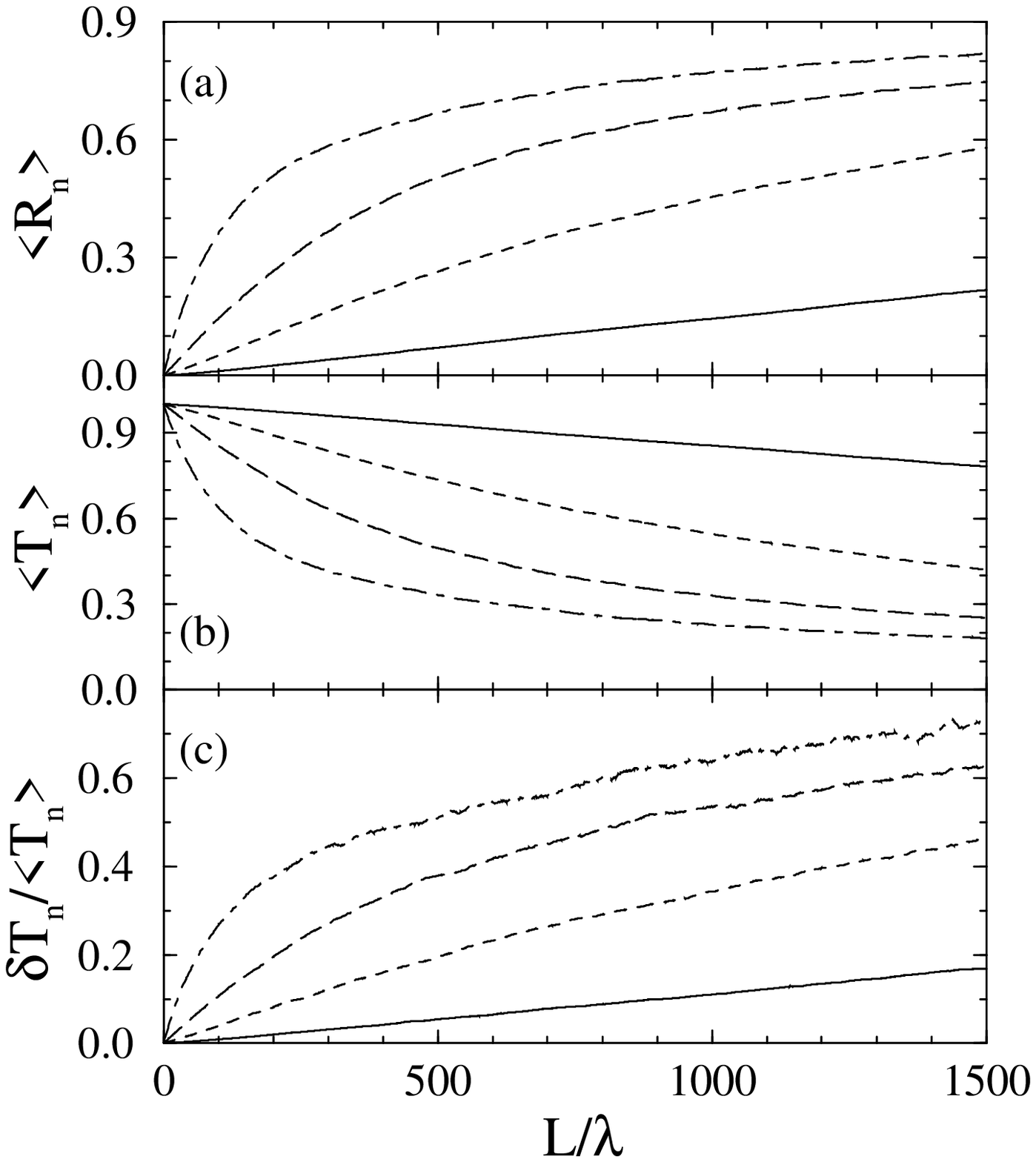}
\caption{Mean total (a) reflection and (b) transmission intensities, and
(c) normalized fluctuations of the total transmission intensities, as 
functions of the length $L$, for the same parameters as in
Fig.~\protect{\ref{fig_tmn}}.}
\label{fig_rtdtn}
\end{figure}

In Fig.~\ref{fig_rtdtn}, we plot the mean total reflection
$\langle R_n\rangle$ and transmission  $\langle T_n\rangle$
coefficients, along with the transmission fluctuations $\delta T_{n}$,
in our 4-mode waveguide with $\delta=0.03\lambda$.
It is evident that these quantities differ substantially from one
incoming mode to another. The larger $n$ is, the larger the mean
reflectance and the smaller the mean transmittance (recall that
energy conservation requires that $R_n+T_n=1$). This could be
qualitatively expected once again, at least in the limit of small
waveguide lengths, in light of the $n$-dependence of the impurity
matrix (\ref{eq_vmn}), which is stronger for incoming modes with larger
transverse momentum $\kappa_n$ (higher $n$). 

\begin{figure}
\epsfxsize=2.5in \epsfbox{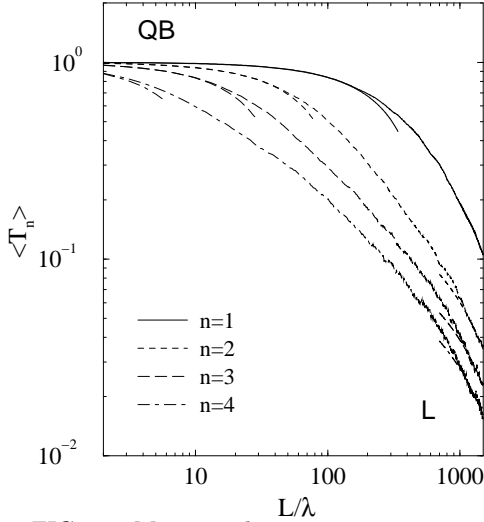}
\caption{Mean total transmission intensities $\langle T_{n}\rangle$ as
functions of the length $L$ in a log-log plot for the same parameters as
in Fig. \protect{\ref{fig_tmn1}}, including fits to the QB and L
regimes.}
\label{fig_tn}
\end{figure}

The overall transport properties of the waveguide for a given incoming
mode $n$ can be understood through the behavior of the mean total
transmission $\langle T_n\rangle$ [see Fig.~\ref{fig_rtdtn}(b)], as 
the summation of $\langle T_{mn}\rangle$ over all outgoing channels
$m$. Figure~\ref{fig_tn}, which shows $\langle T_n\rangle$ in a log plot
for a 4-mode waveguide analogous to that of Fig.~\ref{fig_rtdtn}(b) but
with a rougher surface with $\delta=0.1\lambda$, illustrates this
discussion. To observe a definite transport regime in the total
transmission, either the transport regimes of the different outgoing
modes swap at certain length scales, or one of the $\langle
T_{mn}\rangle$'s predominates over all others. Note that even though
the most transparent mode gives the predominant contribution from a
quantitative standpoint, it is not at all evident that the same is true
for the qualitative behavior (for instance, a steeper, weak decay added
to a larger, but smoother, background would yield as a result a quantity
whose magnitude is of the order of the latter, but whose qualitative
behavior is given by the former weak decay). 
In principle, it can be assessed that transport will obviously
be QB for lengths shorter than $\ell^{QB}_{nn}$, namely, for $L\leq
\ell^{QB}_{nn}$, as Fig.~\ref{fig_tn} reveals through the QB linear
decays (see also Fig.~\ref{fig_tnn}); this can also be verified in 
Fig.~\ref{fig_rtdtn}(b).  Conversely, the exponential decay associated
with L appears beyond waveguide lengths for which the lowest (1n) mode
is localized  $\ell^L_{1n}$, as seen in Fig. \ref{fig_tn} (see also
Figs.~\ref{fig_tmn1} and \ref{fig_tnn}). Finally, unlike the QB and L
regimes, which must always be encountered for sufficiently small and
large lengths, 
respectively, it is not obvious that the $L^{-1}$ dependence (D
regime) in the intermediate region between QB and L transport is
observed. This effect is another manifestation of the entangling of
different transport regimes of the $\langle T_{mn}\rangle$ due to the
combination of surface-type disorder and large mode dispersion. In 
Fig.~\ref{fig_tn}, where the length dependence of $\langle
T_n\rangle$ is shown in a log scale, D should manifest itself through a
linear decay [cf. Eq. (\ref{eq_d})]. It is seen that this decay is
practically absent for most incoming modes. Only within a narrow
waveguide length window for which the D length scales of the
transmission coefficients swap, would the corresponding total
transmission exhibit the expected $L^{-1}$ behavior.  In any case, it is
obvious that the average reflection and transmission coefficients fail
to obey the predicted dependences $\langle T_n\rangle=N^{-1}\langle
g\rangle$ and $\langle R_n\rangle=N^{-1}(N-\langle g\rangle)$ in the
weak localization or D regime \cite{rev,rmt}. 

\begin{figure}
\epsfxsize=2.5in \epsfbox{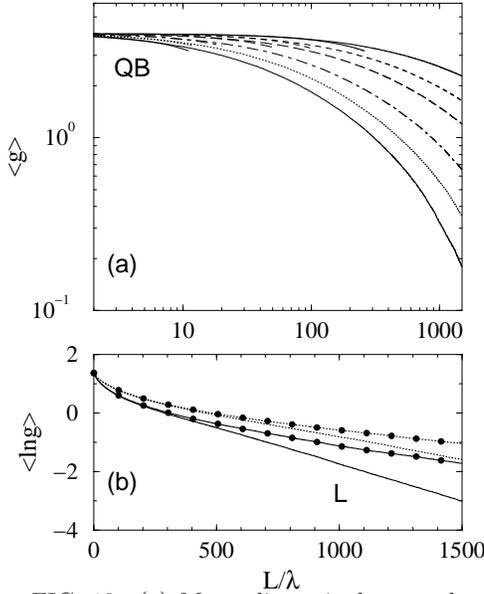}
\caption{(a) Mean dimensionless  conductance as a function of the
length $L$ in a log-log plot for a waveguide width $d/\lambda=2.25$,
supporting 4 modes, with disorder parameters $a/\lambda=0.2$ and
$\delta/\lambda=0.02$ (upper solid curve), 0.03 (dashed curve),0.04
(long dashed curve), 0.06 (dot-dashed curve), 0.08 (dotted curve), and
0.1 (lower solid curve). Averages over $N_p=4000$ realizations.  Fits to
the QB regimes are shown. (b) $\langle\ln g\rangle$ (without dots) and
$\ln\langle g\rangle$ (with dots) for $\delta/\lambda=0.08$ (dotted
curves) and 0.1 (solid curves), revealing  the L regimes. } 
\label{fig_g}
\end{figure}

\begin{figure}
\epsfxsize=2.5in \epsfbox{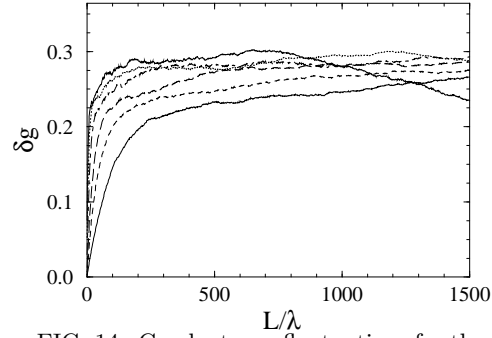}
\caption{Conductance fluctuations for the 
waveguides used in Fig.~\protect{\ref{fig_g}}.}
\label{fig_dg}
\end{figure}

All these transmission properties, stemming from the mixing of QB, D,
and L transports produced by surface disorder, become even more
pronounced in the dimensionless conductance $g$. Figure~\ref{fig_g}(a)
shows a logarithmic plot of $\langle g\rangle$ for 4-mode waveguides
with different surface roughness parameters $\delta/\lambda=$0.02, 0.03,
0.04, 0.06, 0.08, and 0.1, whereas $\langle\ln g\rangle$ is plotted in
Figure~\ref{fig_g}(b) [and also $\ln\langle g\rangle$] for the two
larger $\delta$ values. The corresponding conductance fluctuations are
given in Fig.~\ref{fig_dg}. 
Following the argument mentioned above for the total transmission, now
the QB regime is restricted to the shorter $\ell^{QB}_{nn}$, in this
case $\ell^{QB}_{44}$ [see Fig.~\ref{fig_g}(a)]. This is explicitly
shown in Fig.~\ref{fig_lnn}(a). Likewise, the true L behavior in the
conductance is ensured for lengths beyond which the (11) channel appears
localized; this is seen in Fig.~\ref{fig_g}(b) (in those cases for which
$\ell^L\leq L_{max}$) through the linear decay of $\langle\ln g\rangle$,
and its departure from $\ln\langle g\rangle$ (owing to the transition to
log-normal statistics, for which the dominant contributions arise from
the low-probability realizations that yield large conductances
\cite{frei,nimac}). Recall that, although the actual onset of L thus
appears at slightly different lengths, the localization length for given
roughness parameters as defined from Eq. (\ref{eq_l}) is the same for
all modes ($mn$), and coincides with those for the transmittances and
conductance [see Fig.~\ref{fig_lnn}(b)]. On the other hand, the
absence of the D regime in the conductance curves is explicitly
displayed in Figure~\ref{fig_g}(a). Thus an anomalous conductance
crossover from QB to L regimes is observed for the 4-mode, surface
disordered waveguides with several roughness parameters used in
obtaining the results of Fig.~\ref{fig_g}, giving additional confirmation
of the predictions of Ref. \onlinecite{prl98}. Moreover, the conductance
fluctuations within this anomalous transition regime stabilize in all
cases shown in Fig.~\ref{fig_dg} at a value ($\delta g\approx 0.29$),
which lies below the expected value of the quasi-1D universal
conductance fluctuations (UCF) for a well-defined D regime 
($\delta g\approx 0.364$, cf. Refs. \onlinecite{taan,pedro}).
This lower value of the UCF has been also numerically found in Ref.
\onlinecite{nimac}, but no physical interpretation was given therein.
When entering into the L regime, our results for the waveguides with
rougher surfaces in Fig.~\ref{fig_dg} reveal that the conductance
fluctuations decrease below the UCF region, as expected \cite{taan,nimac}. 

Two comments are in order concerning the anomalous QB-L crossover in the
total transmission and conductance mentioned above. First, it should be
emphasized that surface disorder is not a sufficient condition.
Actually, in the case of surface-disordered waveguides with small mode
dispersion and/or strong inter-mode mixing, so that the D-like regimes
of different outgoing channels coexist, the $L^{-1}$ diffusive
dependence could also be observed. Nevertheless, even if such a D regime
appears, our results still reveal an anomalous behavior, inasmuch as the 
mean total reflection and transmission fail to follow the predicted weak 
localization length dependences, as pointed out above. As a second
remark, it is worth mentioning that the D-like regime is enhanced in the
average resistance (in which contributions from smaller transmission
coefficients predominate), in contrast to the average conductance. The
results presented in Ref. \onlinecite{mnvapl} corroborate these comments,
which thus show no discrepancy with our results.

\section{Conclusions}
\label{sec_con}

The statistical transport properties  of classical waves propagating 
along surface-disordered waveguides have been studied, with special
emphasis on the distinctive imprint introduced by the surface-type
disorder. For this purpose, the invariant
embedding equations for the matrices of reflection and transmission
amplitudes of the guided modes have been obtained. By means of Monte
Carlo simulation calculations, in such a manner that for every surface
realization the corresponding system of coupled differential equations
is numerically solved, the statistical quantities of interest are
calculated. We have focused on the mean reflection and transmission
coefficients, reflectances, transmittances, and conductance, along with
their fluctuations. The interplay between mode conversion and the
scattering processes produced by surface disorder results in new
interesting effects in the physical quantities mentioned.

For a sufficiently long waveguide, transmission into modes
with lower transverse momentum (lower index $m$ according to our
notation) is favored (exhibiting smaller normalized fluctuations), no
matter what the incoming mode $n$ is. The influence of the incoming mode
$n$ is revealed in the decrease of all the transmission coefficients for
higher $n$. For smaller waveguide lengths, we have confirmed, through the
analysis of the length dependence of the mean transmission coefficients
and fluctuations in the case of 4-mode waveguides, the entangling of 
ballistic, diffusive, and localized transmission of modes within the
same waveguide region that was recently reported in
Ref.~\onlinecite{prl98} in the case of 8-mode waveguides.

With regard to the reflection coefficients, enhanced backscattering is
observed when the rough waveguide is long enough, and the enhancement
factor, as defined by the ratio  
$\min[\langle R_{nn}\rangle/\langle R_{mn}\rangle$], can be larger than
2. In fact, the non-diagonal reflection coefficients tend to be smaller
for the reflected modes with lower $m$ for all incoming modes 
$n$. The reflected speckle patterns exhibit reduced fluctuations in
backscattering $(\delta R_{nn})/\langle R_{nn}\rangle\leq 0.5$,
whereas the expected value of 1 is approximately obtained for other
reflected channels ($m\not =n$). Both averages and fluctuations behave
similarly throughout the D and L regimes.

The transmittance, namely, the normalized total energy transmitted for a
given incoming mode $n$, is larger for the lower modes $n$. It should be
noted that, in spite of the small strength of the random component that
is present on one of the waveguide planes, very strong reflectances (of
the order of or larger than 90$\%$) can be observed for sufficiently
long waveguides. This could be relevant in multimode, optical
waveguides with spuriously rough boundaries over long propagation
distances, where it constitutes an unwanted effect \cite{lado}.

We have also analyzed the effect of the entangling of QB, D, and L
transport on the qualitative behavior of the mean transmittance and its
fluctuations, showing an anomalous effective QB-L crossover. This has
also been confirmed in the conductance calculations (average and UCF),
for which the influence of the disorder strength has been shown.

Finally, we would like to mention the very recent works by
Garc{\'\i}a-Mart{\'\i}n {\it et al} on the diffusion-localization transition
\cite{mnvapl} and on the intensity distributions \cite{mnvprl,mnvprlr},
in nanowires with surface-disordered hard-walls consisting of a number
of slices with fixed length and random width, with similarities to the
problem dealt with here. Their numerical results, based on a generalized
scattering matrix formulation exploiting mode matching at each slice,
exhibit also the non-isotropy of the scattering intensities, stressing
however the agreement of the statistics at each transport regime with
the RMT predictions. 

Experimentally, all these effects can be revealed in the transmission
intensities through metal microwave guides, for which our theoretical 
boundary conditions apply very accurately. As pointed out in
Ref. \onlinecite{prl98}, the appropriate geometry would
be a planar waveguide with two metallic plates, one of them at least
randomly rough, with feasible dimensions and roughness parameters (as
derived from the values used throughout this work upon scaling them by
the wavelength in the centimeter range). Similar waveguides but with
tube geometry have been successfully employed in connection with volume
disorder \cite{stoy}. Also in the electromagnetic domain, optical
waveguides or fibers (in the micron range) could be other experimental
devices \cite{lado}, accessible to such measurements, where the
predictions of our calculations can manifest themselves, although in
order to make rigorous quantitative comparisons the boundary conditions
might have to be revised. Furthermore, the propagation of acoustic waves
or other classical waves through confined geometries with appropriate
randomness can be adequately accounted for by our formulation, and thus
similar phenomena might be expected therein. The conductance
calculations can be also of interest in the electronic transport through
nanowires.

\acknowledgments

JASG is grateful to A. Garc{\'\i}a-Mart{\'\i}n, J. A. Torres, J. J. S{\'a}enz, and
M. Nieto-Vesperinas for valuable discussions, and acknowledges support
from the Spanish CSIC, DGES Grant PB97-1221 and CICYT Grant
TIC95-0563-CO5-03.
The work of AAM was supported in part by Army Research Office Grant DAAH
04-96-1-0187. IVY gratefully acknowledges support by EPSRC Grant
GR/K95505.

\begin{figure}
\epsfxsize=2.5in \epsfbox{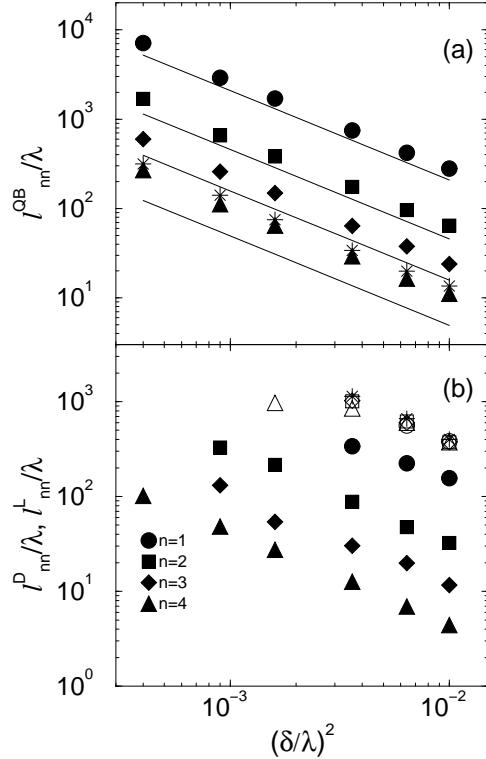}
\caption{Typical decay lengths (in wavelength units) as functions of
the height standard deviation $\delta^2$ (in $\lambda^2$ units), obtained 
from the mean transmission intensities (see text) for $d/\lambda=2.25$
and $a/\lambda=0.2$. Circles: $n=1$; Squares: $n=2$; Diamonds: $n=3$;
Triangles: $n=4$. (a) $\ell^{QB}_{nn}$ from numerical simulation data
(symbols, asterisks denoting the conductance $\ell^{QB}$) and from
perturbation theory (solid lines); (b) $\ell^{D}_{nn}$ (filled symbols)
and $\ell^{L}_{nn}$ (open symbols), and $\ell^{L}$ from the conductance
(asterisks). }
\label{fig_lnn}
\end{figure}

\end{document}